\documentclass[%
 reprint,
 amsmath,amssymb,
 aps,
]{revtex4-1}

\usepackage{graphicx}
\usepackage{dcolumn}
\usepackage{bm}
\usepackage{hyperref}

\hypersetup{
    bookmarks=true,
    unicode=false,
    pdftoolbar=true,
    pdfmenubar=true,
    pdffitwindow=false,
    pdfstartview={FitH},
    pdftitle={Molecular droplets coupled to intense cavity fields},
    pdfauthor={Stefan Kuhn},
    pdfnewwindow=true,
    colorlinks=true,
    linkcolor=red,
    citecolor=blue,
    filecolor=magenta,
    urlcolor=cyan   
}

\begin{document}

\preprint{APS/123-QED}

\title{Cavity cooling of free silicon nanoparticles in high-vacuum}

\author{Peter Asenbaum}
\author{Stefan Kuhn}
\author{Stefan Nimmrichter}
\author{Ugur Sezer}
\author{Markus Arndt}

\affiliation{University of Vienna, Faculty of Physics, VCQ, Boltzmanngasse 5, A-1090 Vienna, Austria}

\date{\today}

\begin{abstract}
Laser cooling has given a boost to atomic physics throughout the last thirty years since it allows one to prepare atoms in motional states which can only be described by quantum mechanics. Most methods, such as Doppler cooling\cite{Hansch1975}, polarization gradient cooling\cite{Dalibard1989} or sub-recoil laser cooling\cite{Kasevich1992,Lawall1994} rely, however, on a near-resonant and cyclic coupling between laser light and well-defined internal states. Although this feat has recently even been achieved for diatomic molecules\cite{Shuman2010}, it is very hard for mesoscopic particles.  It has been proposed that an external cavity may compensate for the lack of internal cycling transitions in dielectric objects\cite{Horak1997,Vuletic2000} and it may thus provide assistance in the cooling of their centre of mass state. 
Here, we demonstrate cavity cooling of the transverse kinetic energy of silicon nanoparticles propagating in genuine high-vacuum ($< 10^{-8}$ mbar). We create and launch them with longitudinal velocities even down to $v \leq
 1$ m/s using laser induced thermo-mechanical stress on a pristine silicon wafer. The interaction with the light of a high-finesse infrared cavity reduces their transverse kinetic energy by more than a factor of $30$.  This is an important step towards new tests of recent proposals to explore the still speculative non-linearities of quantum mechanics\cite{Bassi2013,Penrose1996,Nimmrichter2011,Adler2009}  with objects in the mass range between $10^7$ and $10^{10}$ amu. 
\end{abstract}

\maketitle

The original idea of cavity cooling\cite{Horak1997,Vuletic2000} has already been realized with atoms \cite{Maunz2004}, ions\cite{Leibrandt2009} and atomic ensembles\cite{Chan2003}, even to below the recoil limit\cite{Wolke2012}.  Such experiments benefit from advanced preparation methods and resonant optical forces which are available for atoms. In contrast to that, the manipulation of mesoscopic particles cannot profit from any internal resonance. On the other hand, non-resonant light forces are non-satiating in nature and one can benefit from high optical intensities for cooling and high scattering rates to gain detailed insight into the cavity cooling process. 
Several groups have started to load silica (SiO$_2$) micro- or nanoparticles into buffer-gas assisted optical dipole traps to successfully subject them to optical cooling\cite{Li2011,Kiesel2013,Gieseler2012}. In contrast to that, we here demonstrate efficient transverse cavity cooling of pure and freely propagating silicon (Si) nanoparticles in a true high-vacuum environment.

\begin{figure}[h!]
\centering
		\includegraphics{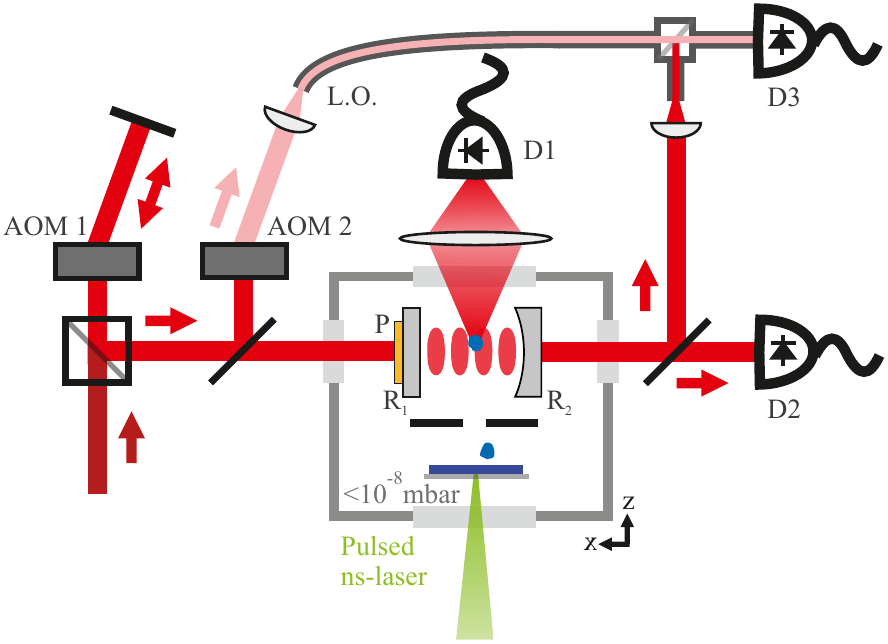}
\caption{\textbf{Cavity cooling of pure silicon nanoparticles in a high-vacuum environment.}
The fibre laser beam (1560 nm) is double-passed in AOM1 ($2\times40$ MHz) to stabilize the laser frequency on short timescales to the cavity. Over long timescales the piezo-electric transducer P locks the cavity to the laser frequency.  A part of the stabilized beam is split off and frequency-shifted by AOM2 to serve as the local oscillator beam (LO). The other part (1 mW) is coupled into the high-finesse cavity inside a high vacuum chamber ($<10^{-8}$ mbar). The cavity has a finesse of $3\times 10^5$, a waist of $w=65 \mu$m at the centre and consists of one flat and one curved mirror ($R2=25$mm). The transmitted intensity, which is a direct measure for the intra-cavity power, is detected by photodiode D2. A part of the transmitted intensity is fibre-coupled, overlapped with the local oscillator beam and detected by D3. The phase of the resulting beating signal allows us to extract information about the cavity phase. A pulsed ns-laser (532 nm, 15 mJ) is focused onto a 0.5 mm thick silicon wafer to create and launch slow silicon nanoparticles. The wafer is attached to the top of a quartz plate. To avoid contamination of the cavity mirrors the particles travel through a 1 mm aperture before they enter the cavity mode. Light scattered from within the cavity is focused onto the photodiode D1. }
\label{figure1}
\end{figure}

The outline of our experiment is shown in figure \ref{figure1}:  We generate pure silicon nanoparticles in situ by directing a pulsed focused laser beam onto the back-side of a clean silicon wafer. The particles emerging from the front-side travel upwards against gravity and interact with the standing light wave inside the high-finesse cavity. 
A fibre laser operating at $\lambda = 1560$ nm seeds the fundamental TEM$_{00}$ cavity mode. Its frequency is modulated via the double-pass acousto-optic modulator AOM1 and the locking signal is derived from the cavity mirror's birefringence\cite{Asenbaum2011}. A part of the modulated laser light is split off, frequency-modulated by 40 MHz in AOM2 and mixed with the light that is transmitted through the resonator on photodiode D3. The resulting 40 MHz beating signal is digitized to extract the phase difference between the two beams.

In parallel, the time evolution of the intra-cavity power is monitored via the intensity that is captured on D2 behind the cavity mirror R2. The photodiode D1, positioned under right angle to both the cavity axis and the polarization of the incident laser beam, is used to record the scattered light when a silicon particle passes the cavity.  The detector is dominantly sensitive to the dipole radiation pattern, which reveals details about the coupling strength between the transiting particle and the cavity field.

An object of polarizability $\alpha$ experiences the optical potential $V_{opt}=-\hbar U_0 |a|^2 f^2 (x)$, where $f(x)$ describes the intracavity mode function and $a$ the field amplitude. The particle effectively increases the cavity length due to its index of refraction and it shifts the cavity resonance by $U_0 f^2 (x)$,  where $U_0=\alpha \omega_L /2\epsilon_0 V$ , $V$ the mode volume and  $\omega_L$  the laser angular frequency\cite{Nimmrichter2010}. Due to the cavity's high finesse, even a small length change leads to a significant phase shift of the cavity field. If the incident laser beam of frequency $\omega_L$ is red-shifted with respect to the cavity resonance $\omega_C$ ($\omega_L-\omega_C<0$) any normal-dispersive particle (high-field seeker, index of refraction $n>1$) will tune the cavity resonance closer to the laser frequency and increase the intra-cavity power.

When a particle moves along the cavity mode it creates a time-dependent phase shift and intensity modulation. The intensity is maximal when the particle leaves an antinode of the optical lattice and minimal when it moves towards it. Therefore, the particle gains less energy when it runs downhill than it loses while running uphill. This Sisyphus-type cooling \cite{Horak1997} is most efficient when the field maxima coincide with the steepest slope of the coupling curve, which is satisfied for particles with a Doppler shift of $kv\approx \Delta$. In order to achieve the maximum intensity change per phase change, the cavity is detuned by one resonator line width $\kappa$ to $\Delta\approx-\kappa$.

\begin{figure}[h]
	\centering
		\includegraphics[width=0.4\textwidth]{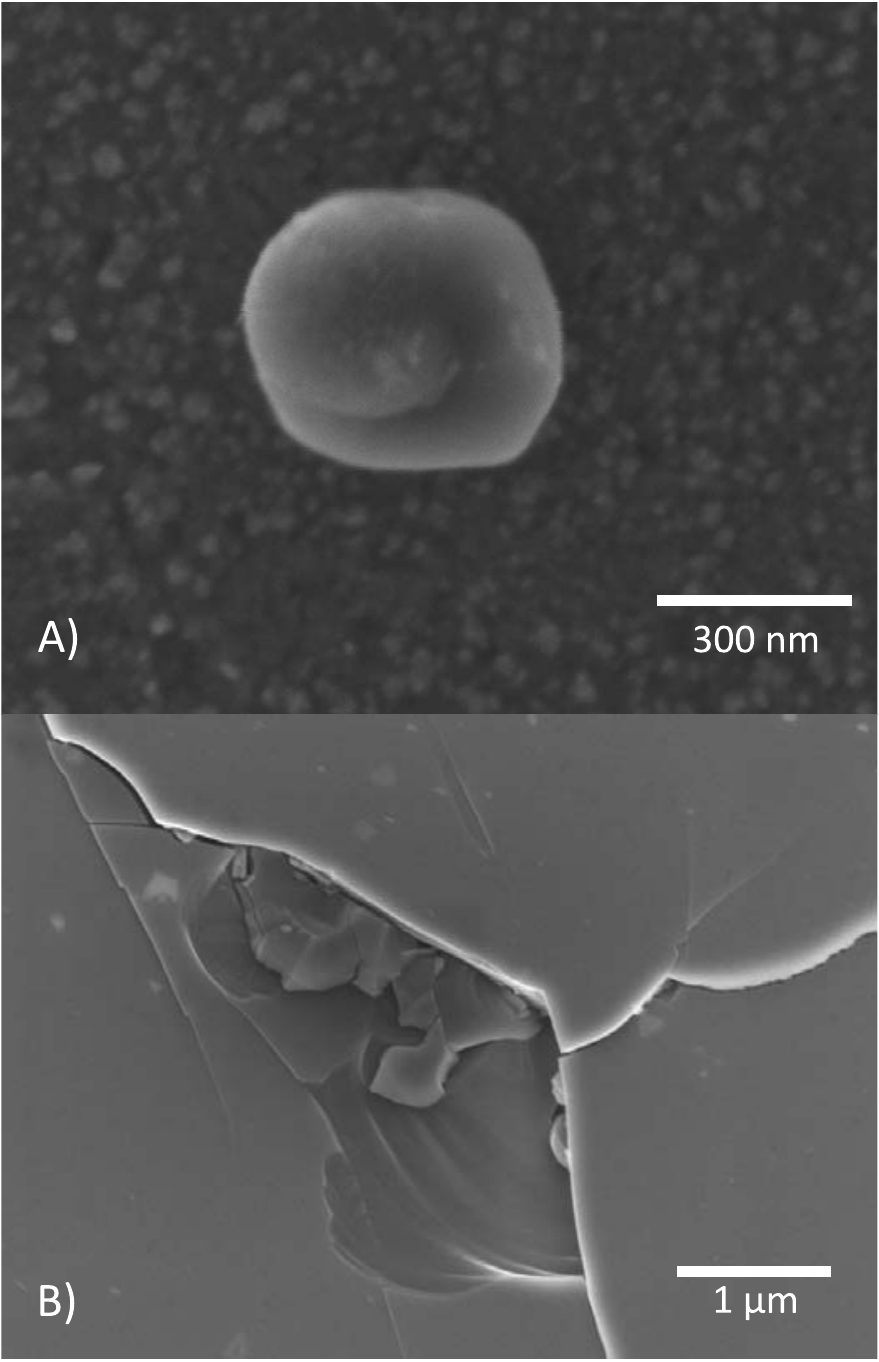}
\caption{ A) A scanning electron micrograph (SEM) shows a typical silicon nanoparticle that is released by laser induced thermo-mechanical stress(LITHMOS) ablation from a pure silicon (111) surface. Particles in the size range from below 100 nm to beyond 1000 nm have been observed, with a variation of shapes. While a full model of the release mechanism is still subject to future research, the cracks and defects on the front-side of the back-irradiated silicon wafer (B) corroborate the hypothesis that micro cracks are the emission sites of nanosilicon. Energy dispersive x-ray spectroscopy (EDX) on the ejected nanomaterial allows us to identify silicon as the principal element, with oxygen as an occasional contaminant in some particles.}
\label{figure2}
\end{figure}

One of the grand questions in nanoparticle cavity cooling, in comparison to the case of atoms, is how to prepare slow neutral particles in the first place.  We use laser-induced thermo-mechanical stress (Suppl. Inf. 1) on a pristine surface to generate and release clean silicon nanoparticles under excellent vacuum conditions. The scanning electron micrograph (SEM) of figure \ref{figure2}A shows a prototypical nanoparticle that was released from the front-side of a silicon wafer when its back-side was irradiated by pulsed laser light. This method generates objects of varying diameters and shapes, ranging from spheres of less than 100 nm across to arbitrarily shaped fragments well beyond one micrometre. An SEM image of the wafer's front surface (figure \ref{figure2}B) corroborates the proposed mechanism of particle formation: Microscopic cracks appear as a result of the laser-induced stress\cite{Du2012}. Here we post-select the smaller fraction of particles since objects with a radius above 200 nm will average over the standing wave potential and see reduced cooling forces\cite{Pflanzer2012}.

\begin{figure*}
\includegraphics{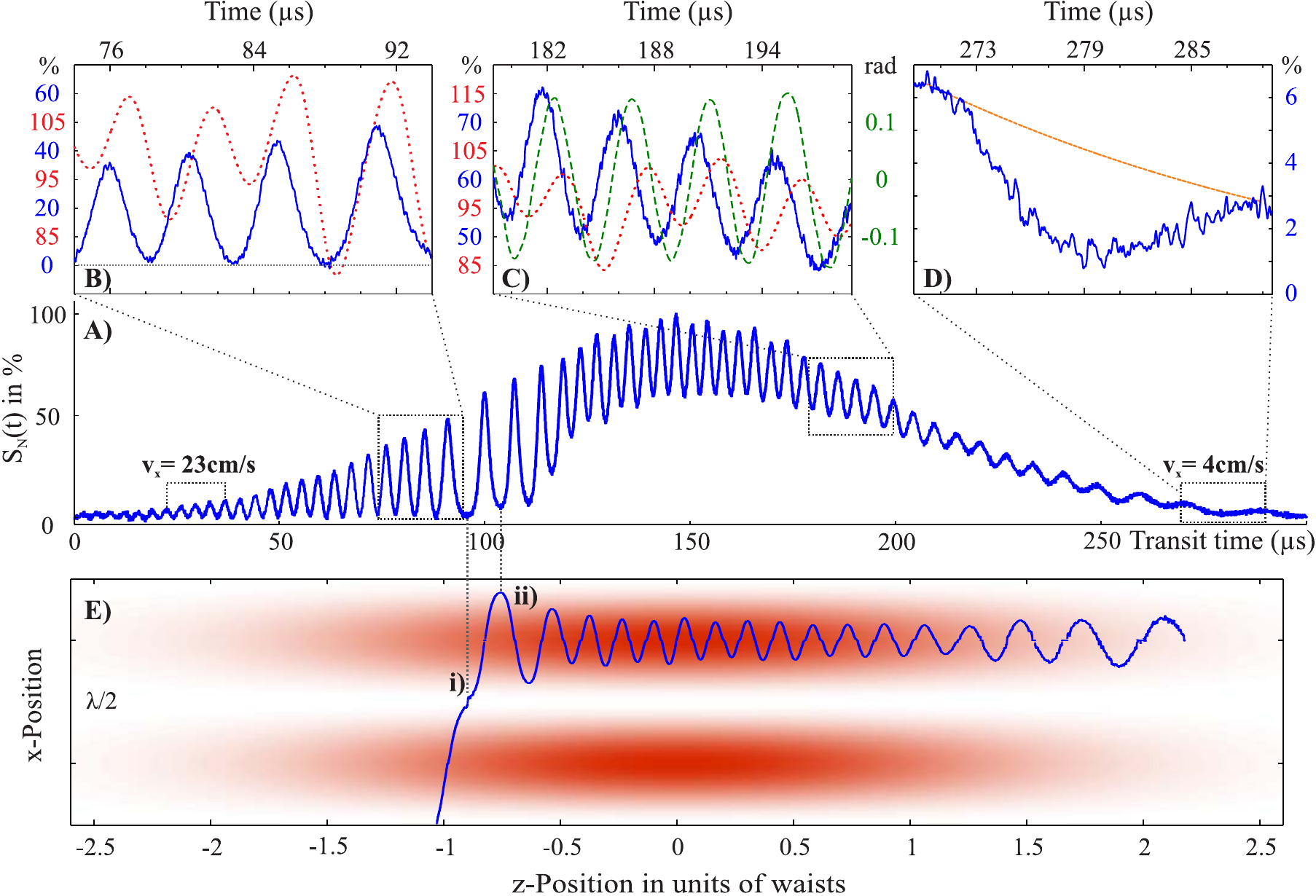}
\caption{\textbf{A)} Cooling of a silicon nanoparticle traced by its scattered light in transit through the high-finesse IR cavity.
	We plot the normalised scattering $S_N (t)$ as a function of the particle's transit time through the cavity mode.  This serves us to extract the particle's entrance and exit velocity along the cavity axis (panel D), as well as the particle trajectory (panel E).  The modulation in the Gaussian envelope (panel B) is more pronounced in the entry wing than in the exit wing. This temporal asymmetry is indicative of cavity cooling which reduces the velocity along the axis from 23 cm/s to 4 cm/s.
	\textbf{B)}When the particle enters the cavity mode it runs over the standing wave and creates a sinusoidal modulation in $S_N (t)$ with $100\%$ visibility (blue line). The cavity intensity (red dotted line) is modulated by the particle, however delayed due the finite cavity response time. The intensity maxima coincide with the turning points of $S_N (t)$, which corresponds to optimal cooling. 
	\textbf{C)}The particle is trapped and oscillates around the antinode. The cavity phase (green, dashed) is modulated by the particle with a slight time delay and the intensity (red, dotted) is again time delayed. Although the particle is trapped, it is still further cooled since the intensity is still close to the maximal slope of $S_N (t)$. When the particle leaves the centre of Gaussian mode, the trapping frequency decreases again.
\textbf{	D)}The evaluation of the particle transit time through the last notable oscillation of the normalized scattering curve in the exit wing, in comparison to one of the first wiggles in the entrance wing, allows us to quantify the cavity cooling.  
\textbf{	E)} Qualitative reconstruction of the particle's trajectory in two spatial dimensions (Suppl. Inf. 6). 
The last free transit through a node of the standing light wave (i) and the first onset of trapping by the optical potential (ii) are correlated with the normalised scattering $S_N (t)$ of panel (A) at the times that are indicated by the dotted lines.}
\label{figure3}
\end{figure*}

The light in photodiode D2 measures the intra-cavity power $I_c (t)$ and therefore also the expected optical forces. The intensity $I_s (x(t))$ scattered into photodiode D1 is a measure for the light intensity at the particle position.  We derive a normalised scattering coefficient  $S_N (t) \equiv(I_s/I_c )/$max$(I_s/I_c)$ and plot it in figure \ref{figure3}A. It is an intensity-independent measure for the scattering probability which contains the information about the cooling process. It relates to the cavity mode function via $S_N (t)=\sin^2(kx(t))  \exp(-2 z^2 (t)/w^2 )$,  where $w$ is the mode waist. 

We can extract the particle position $x(t)$ from $S_N (t)$  as described in Suppl. Inf. 3 and 6. When the particle passes the cavity with the velocity $v_z$, we expect a Gaussian envelope of $S_N (t)$ with an 
$e^{-1/2}$ half-width of $w/v_z$.  From this we deduce a z-velocity of $v_z=0.7$ m/s for figure \ref{figure3}A.

In the beginning of the Gaussian envelope the particle is almost unaffected by the weak optical potential and it travels under a small angle to the vertical direction. One expects a sinusoidal modulation of $S_N (t)$ with a period of $T=\lambda/2v_x$, where $v_x$ is the initial velocity along the cavity axis. A maximum in $S_N (t)$ reveals the particle's passage through an antinode of the standing wave. When the nanosilicon approaches the cavity centre along the z-axis, its transverse velocity $v_x$ is repeatedly reduced by the desired intra-cavity Sisyphus effect. When the particle's kinetic energy has fallen below the optical potential, the silicon ball is channelled vertically between two standing-wave nodes. The first velocity inversion, i.e. the onset of channelling, can be identified with a first local (non-zero) minimum of $S_N (t)$. It is possible to reconstruct the particle trajectory with good reliability \cite{Mabuchi1996} to illustrate cooling and temporary channelling, as shown in figure \ref{figure3}E.

The oscillation amplitude of the trapped particle is a measure for its kinetic energy and the cooling force is still active since an intensity maximum still coincides with the steepest slope of $S_N (t)$. However, the energy loss per cycle decreases since the particle now remains bound to regions of similar coupling. The overall energy loss of the nanosilicon leads to a clearly visible asymmetry in $S_N (t)$. The particle enters the cavity with a horizontal velocity of $v_x=23$ cm/s and leaves it with $v_x=4.0(2)$ cm/s. This corresponds to a reduction in the transverse kinetic energy by a factor of more than 30 (Suppl. Inf. 5).

While the particle moves along the standing-wave it modulates the cavity phase (figure \ref{figure3}C). From the amplitude of this modulation we can extract an effective coupling frequency $U_x=2.3(4)\kappa$. Due to an uncertainty in the y-position, field averaging caused by the particle's finite size and possibly an anisotropic shape, $U_x$ is a lower bound for the maximal coupling frequency $U_0$. In order to estimate $U_0$ and to derive a particle size we measure the trapping frequency $f_{trap}$, while the particle is channelled. We find (Suppl. Inf. 4) that the measured trapping frequency of $f_{trap} = 145$ kHz corresponds to a silicon sphere with a radius of about $150$ nm  ($2 \times 10^{10}$ amu, polarizability of $4\pi \epsilon_0\times 2.7 \times 10^9 \AA^3$, $N= 7\times 10^8$ atoms) which misses the cavity centre by about 0.4 waists.

Summarizing, we see substantial cavity cooling of pure silicon (Si) nanoparticles in a high-vacuum environment and the cavity-assisted readout also provides information about the particle's polarizability, mass and velocities.
Our demonstration is an important step towards quantum experiments in the regime of high masses. While first new bounds on continuous spontaneous localization\cite{Nimmrichter2011,Haslinger2013} will already come into reach with quantum interferometry at a mass range of $10^5$ amu, certain tests will require nanoparticles of $10^8-10^{10}$ amu or more\cite{Kaltenbaek2012}. Objects that big  are, in principle, still compatible with an interferometer concept that prepares and probes quantum coherence using pulsed optical ionization gratings\cite{Reiger2006}, especially when working in the time domain (OTIMA)\cite{Haslinger2013}. The work function of pure silicon (Si) is well matched to it.

Future quantum experiments will still require transit and coherence time of a few seconds, at least. Our successful demonstration of cavity cooling by a factor of 30 in energy is a promising starting point. It is interesting to consider multimode cavity cooling \cite{Habraken2013} to reduce trapping and to increase the cooling rate. Experiments are also on the way to extend the source to cold and mass-selected single particles on demand.

\subsection*{Acknowledgments}
We are grateful for financial support by the Austrian Science Funds (FWF) in the projects DK-CoQuS (W1210-2) and Wittgenstein (Z149-N16) as well as the ESA project No. 4000105799/12/NL/Cbi  as well as the EC program Nanoquestfit (304886).  We acknowledge support by Stephan Puchegger and the faculty centre for nanostructure research at the University of Vienna in imaging the silicon nanoparticles.  We are grateful for fruitful discussions with Helmut Ritsch, Claudiu Genes, Wolfgang Lechner, Jack Harris, Markus Aspelmeyer, Philip Schmid and Clemens Mangler. 

\bibliographystyle{unsrt}
\bibliography{graphy}
\newpage

\begin{appendix}
\section*{Supplementary Information}
\subsection{Creation of a silicon nanoparticle beam in a high vacuum environment}

 \textit{Ablation by laser-induced thermo-mechanical stress (LITHMOS).}  
Green light from a frequency doubled Nd:YAG laser (Innolas, 532 nm, 15 mJ, 4 ns) is directed onto the back-side of a $500 \mu$m thick and chemically cleaned silicon wafer (111 cut) which is attached to a quartz plate and suspended in high-vacuum ($< 1\times10^{-8}$\, mbar). The origin of this method is similar to that of laser-induced acoustic desorption (LIAD) \cite{Zinovev2011,Dow2012} with the difference that LIAD experiments are targeted at releasing surface adsorbents. Here, we release nanomaterial from the pristine silicon surface itself. The purity of the material is verified with energy dispersive X-ray scattering (EDX) inside a scanning electron microscope (SEM).

The particles are launched by the laser, but not necessarily in the same laser pulse that creates them. We also observe the formation of silicon dust on the front-surface which is then cleaned in subsequent laser shots. Silicon nanoparticles offer promising properties for cavity cooling as well as for future quantum interference studies, alike. They exhibit a very high polarizability-to-mass ratio in comparison to many other dielectrics and are therefore well susceptible to optomechanical forces. Their high refractive index, $n (1560 $nm$)= 3.47(1)$, is associated with a relative permittivity of $\epsilon_R=n^2=12.04(1)$. Using the Clausus-Mosotti relation we can relate the atomic polarizability and bulk permittivity and find a value of $\alpha=4\pi \epsilon_0  R^3 (\epsilon_R-1)/(\epsilon_R+2)$.
The work function of silicon is $W=4.52$ eV and indicates that single-photon ionization can easily be implemented using a UV or VUV laser, as required for future OTIMA interference experiments\cite{Reiger2006,Haslinger2013}. 

\subsection{Laser-cavity stabilization} 
We use an IPG photonics fibre laser ELR-10-1560-LP-SF at 1560 nm. It has a short term (50 ms) line-width of 20 kHz. Laser and cavity are locked by a slow and a fast feedback loop. Fast feedback is implemented via frequency modulation of the incident laser light in the double-pass AOM 1 (figure \ref{figure1}). It operates with unity gain at 150 kHz. A slow active detuning and feedback loop is implemented by a piezoelectric displacement of the flat cavity mirror R1. This loop has a bandwidth of about 10 kHz. We exploit the intrinsic mirror birefringence to generate a dispersive locking signal without the need for actively modulating any optical element\cite{Asenbaum2011}.
The high-finesse cavity is not only sensitive to the transiting nanoparticles but also to the tiny thermal expansion that is caused by the residual absorption in the mirror coating\cite{Carmon2004,Dube1996}. A rise in the intra-cavity intensity will lead to an even increased heating of the mirror coating and further thermal expansion. When the laser is red-detuned with respect to the optical cavity, this can result in self-stabilization but also to self-driven length oscillations with frequencies of the order of a few kHz in case of high laser intensities. In order to avoid these instabilities, the feedback modulation of the laser frequency must be fast enough. This fast feedback suppresses the slow cavity phase shift that is related to the particle's passage through the Gaussian envelope along z. Only the oscillatory coupling along x, caused by the particle's motion across the intra-cavity standing light wave, remains. 

	\subsection{Time evolution of the cavity field}
The time evolution of the cavity field amplitude $a(t)$ can be written as 
\begin{align}
	\dot{a}(t)=\eta- [\kappa-i\Delta-iU_0 f^2 (x(t))]a(t) ,
\end{align}
where $\eta$ is the pump field, $\Delta$ the laser-cavity detuning, $\kappa$ the cavity loss rate (line width), $U_0$  the cavity shift and $f(x)$ the cavity mode function, which is explored by the particle moving along $x(t)$.
A formal integration of the cavity equation from time $t=-\infty$  yields
\begin{align}
&a(t)= \nonumber \\ 
&\eta \int_{-\infty}^{t} dt' \exp \left(-(\kappa-i \Delta) (t-t') + i U_{0} \int_{t'}^{t} dt'' f^{2} (\textbf{x}(t'')) \right)
\end{align}

In the absence of any particle the cavity assumes the stationary state $a_0=\eta/(\kappa-i \Delta)$.  To lowest order, a slowly moving particle would lead to a quasi-stationary field amplitude $a(t)\approx  \eta/(\kappa-i(\Delta+U_0 f^2 (x(t))) )$.   The formal solution describes the delayed reaction of the cavity to the moving sphere, which primarily modulates the phase of the field. Since the cavity reaction time scale $1/\kappa$ is finite, the particle-induced phase shift turns into an intensity modulation only after another delay. This explains the hierarchy of the observed phenomena in the cavity: The scattered light is an immediate signature of the particle position, followed by the delayed phase reaction of the cavity, and its intensity change, which reacts the slowest.

\subsection{Optical forces on a dielectric particle in a cavity mode}
\begin{figure}[h]
	\centering
	\includegraphics[width=0.45\textwidth]{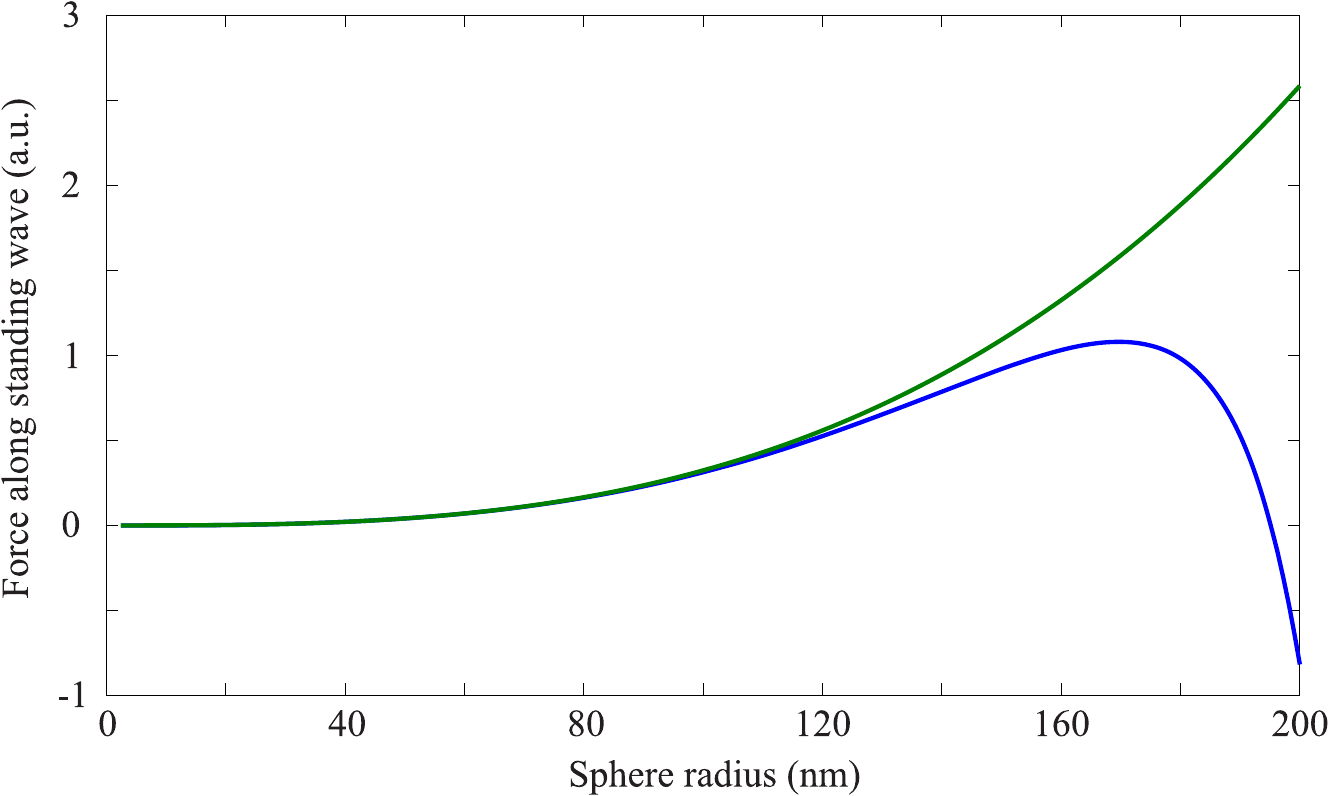}
\caption{Maximum force along the cavity axis for a sphere as a function of its radius R in case of a point-like treatment ($F_x \propto U_0$, green), and a full description ($F_x \propto U_x$, blue), which takes the finite particle size into account.}
\label{figureS1}
\end{figure}

The optical properties of the silicon particle in the presence of a standing-wave cavity field $E_0$  can be described using Mie scattering theory\cite{Hulst1957,Bohren1983}. The internal and scattering field components, $E_{int}$ and $E_{sca}$, can be expressed in terms of a spherical wave expansion. The light-induced force acting on the sphere is then computed by integrating the Maxwellian stress tensor that is associated with the modified external field components $E_{ext}=E_0+E_{sca}$ and $B_{ext}$ over the sphere surface\cite{Jackson1999}. Apart from weak transverse forces due to the Gaussian profile of the cavity field, we find that the force $F=F_x e_x$ is oriented along the standing-wave axis, and we may express it in terms of the coupling frequency $U_x$ as   $F_x=-\hbar kU_x |a|^2  \sin(2kx) \exp(-2(y^2+z^2 )/w^2 )$.
Figure \ref{figureS1} shows the transverse force for a point-like particle (green) compared to a full description (blue) which takes the finite particle size into account. For a sphere radius over R=120 nm the force along the standing wave, starts to deviate from the point particle approximation with the effective coupling frequency $U_0=(2 \pi \omega_L R^3)/V  (\epsilon-1)/(\epsilon+2)$. At roughly 190 nm the force vanishes and changes its sign for even larger radii.

The above force determines\cite{Salzburger2009} the equation of motion along x,
\begin{align}
\ddot{x}(t)=-\frac{\hbar k}{m} U_x |a|^2  \exp[-2z^2 (t)/w^2 ]  \sin(2kx(t)),
\end{align}

where we assume that the particle traverses the cavity mode along the z-axis. From this we extract 
the harmonic frequency $f_{trap}$ for a particle trapped at an antinode of the standing wave using
$f_{trap}=\frac{1}{2 \pi}\sqrt{\frac{k P_{in}}{m c}  \frac{U_x}{\kappa}}$ , where $P_{in}$  is the incoupled power, $k=2\pi /\lambda$  and $c$  the speed of light.
We measure a frequency of $138$ kHz at a normalized scattering $S_N=0.9$ where the oscillation amplitude explores 33\% of the trap potential. Taking the anharmonicity\cite{Johannessen2010} into account we deduce a harmonic trap frequency $f_{trap}= 145$ kHz. For a silicon point particle maximally coupled to the cavity mode one expects $f_{trap}=183$\, kHz. The difference is attributed to the non-maximal coupling of a finite-sized particle. We measure $U_x=2.3(4)\kappa$, which corresponds to the passage of a silicon sphere with a radius of about 150 nm that misses the cavity centre by 0.4 beam waists $w_y$ along the y-axis. 
The mass of a silicon sphere with a radius of 150 nm amounts to  $m_{r=150nm}= 2\times10^{10}$  amu, with $\rho=2.33(1)$ g/cm$^3$ the bulk mass density and $m(Si)=28.086(1)$amu.  This corresponds to about $7\times10^8$  silicon atoms in the ball and it is in the range of explorations that have been proposed in the quest for non-linear extensions of quantum mechanics\cite{Nimmrichter2013,Nimmrichter2011,Kaltenbaek2012}. 
\subsection{Velocity of the particle in the optical potential}

Due to cavity cooling, the velocity of the particle along $x$ is reduced to a level at which it is influenced by even a small field in the exit wing of the Gaussian cavity beam. The corrugated optical potential modulates the velocity of the particle. At a maximum of the coupling, for instance, it is higher than outside the cavity mode. 
Assuming a static field amplitude and a constant velocity along $z$, the moving particle experiences the time dependent potential $V_{opt} (t)=-\hbar U_x \left|a\right|^2 \cos^{2}(kx(t))  \exp(-2v_z^2 t^2/w^2 )$ and the force
\begin{align}
	F_x (t)=-\hbar U_x \left|a\right|^2 k \sin(2kx(t))  \exp(-2v_z^2 t^2/w^2 ) .
\end{align} 
One can formally integrate the equation of motion to obtain
\begin{align}
	v(t)&=v_0+\int_{-\infty}^{t} dt'  F_x (t')/m \nonumber  \\
	&= v_0-\frac{U_x |a|^2 k}{m} \int_{-\infty}^{t} dt' \sin2kx(t') \exp \frac{2 v_{z}^{2} t'^2}{w^2}
\end{align}

For a weak potential this can be approximated to first order by setting $x(t)=v_{0} t$,

\begin{align}
	v(t) & \approx v_0-\frac{\hbar U_x |a|^2 k}{m} \int_{-\infty}^{t} dt'   \sin(2k v_0 t' )  \exp(-2v_z^2 t'^{2}/w^2 ) \nonumber \\
				&=v_0+\frac{\pi}{2} \frac{kw \hbar U_x |a|^2}{2m v_{z}}  \exp(-k^2 w^2 v_0^2/2v_z^2 )\nonumber \\
				& \times \mathrm{Im}\left( \mathrm{erf}\left[\frac{\sqrt{2} v_{z} t}{w}+i \frac{kw v_{0}}{\sqrt{2} v_z}\right] \right).
\end{align}

\begin{figure}[h]
	\centering
		\includegraphics[width=0.45\textwidth]{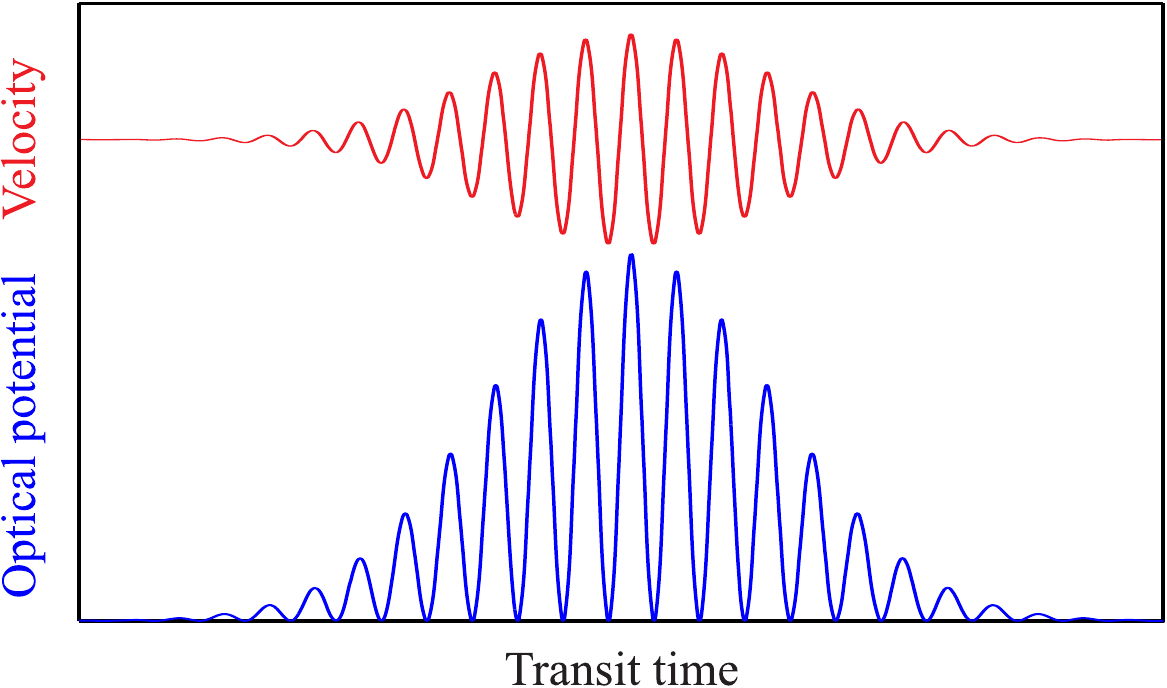}
\caption{Calculated optical potential and velocity for a test particle which runs over a weak standing wave with a static field amplitude}
\label{figureS2}
\end{figure}

\begin{figure}[h]
	\centering
		\includegraphics[width=0.45\textwidth]{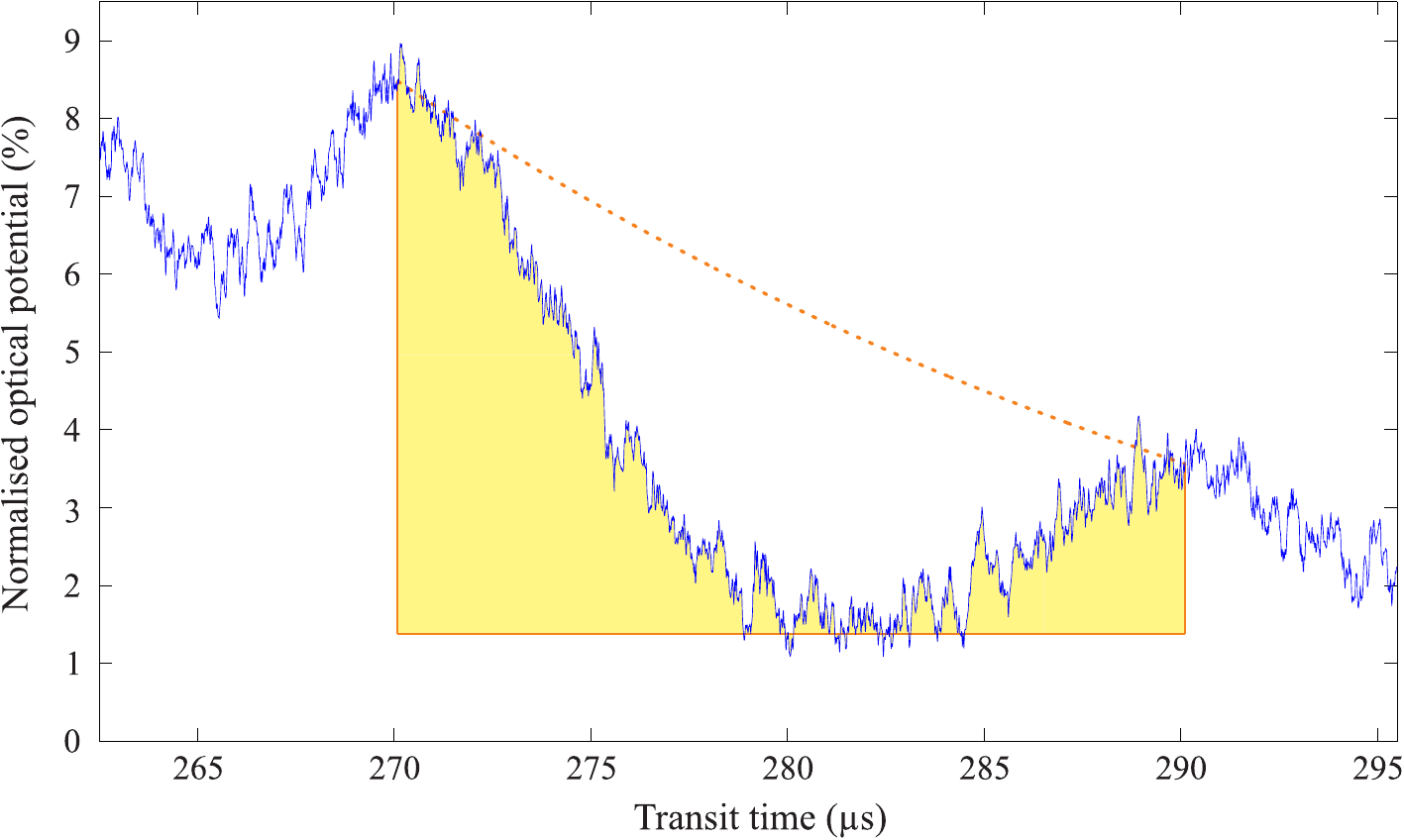}
\caption{A measurement of the area underneath $I_S$ (particle in figure 3) and under the Gaussian envelope can be used to estimate how accurately the velocity in the (weak) exit potential reflects the velocity of the finally untrapped particle.}
\label{figureS3}
\end{figure}

In figure \ref{figureS2} this expression (red) is plotted for a test particle together with its optical potential (blue). At an antinode/node of the field the velocity is higher/lower than far out of the cavity mode. It oscillates around its final exit value $v_0$.
The particle velocity at finite coupling will be the same as far beyond the mode as long as the time spent at high and low potential averages out. In order to determine whether this is the case we integrate the scattering intensity  $I_S=I_C S_N (t)\propto V_{opt} (t)$ between the last two maxima over time, subtract the offset (figure \ref{figureS3} yellow area, $A_1$) and compare it with the area $A_2$ under the Gaussian envelope (area enclosed by the orange lines). If a particle spends more time at low fields, this will result in a smaller area under $I_S$.
For the numerical example of figure \ref{figureS2}, the 'measured velocity' $v_m$ at finite field and the 'true final exit velocity' $v_f$ at zero field will be equal and we find an area ratio of $r_A \equiv (2A_1)/A_2 =1$. 
In the parameter regime $(w/v_z \gg \lambda/2v_x)$  the Gaussian envelope causes only a negligible deviation of $r_A$ from 1.

\begin{figure}[h]
	\centering
		\includegraphics[width=0.45\textwidth]{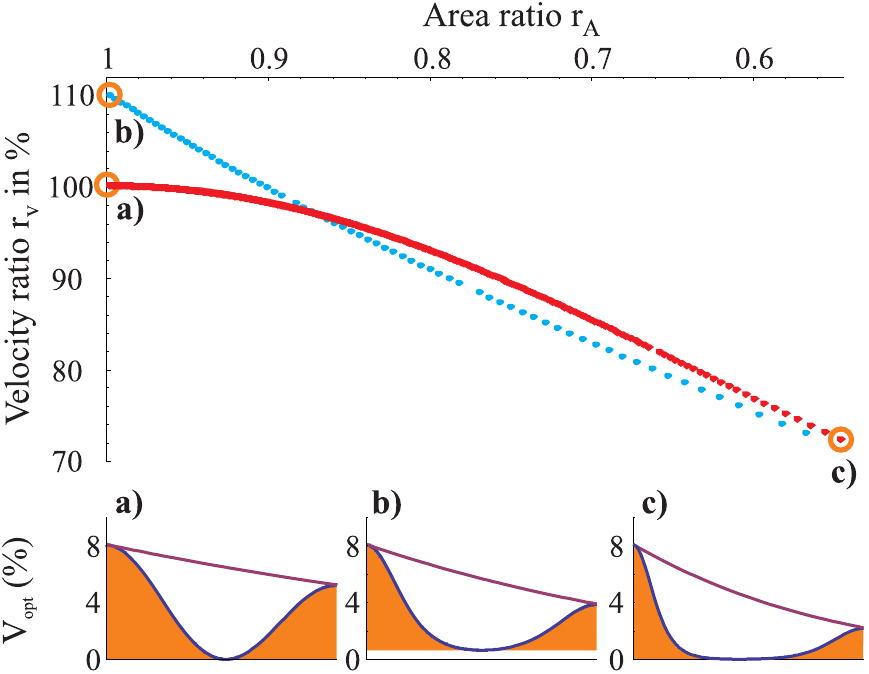}
\caption{Discrepancy in velocity estimate $r_v$  (in $\%$) as a function of $r_A$ for trapped (blue) and un-trapped (red) trajectories. The panels a-c) show the three extremal points of the graph: A freely running (a), a trapped (b) and a nearly trapped (c) particle.
a) The particle is hardly influenced by the optical potential while it moves over the standing wave. Therefore, both $r_v$ and $r_A$  are close to 1.
b)The particle is trapped in the standing wave and $r_A\approx 1$. The measured velocity is about 10\% higher than the actual final velocity of the particle after leaving the cavity mode.
c) A particle at the border between being trapped and untrapped. The measured velocity appears to be slower than it actually is after leaving the cavity (28\%). It also shows a deviation of $r_A$  from 1 by more than 45\%.}
\label{figureS4}
\end{figure}

In contrast to that, a particle that is nearly or weakly trapped will spend more time in the low field region than in the high field. This results in $r_A<1$ and we find that we would underestimate the true velocity  $r_v\equiv v_m/v_f <1$.
On the other hand if we overestimate the true velocity in our measurement we obtain $r_v>1$. By simulating trajectories for different field strengths we examine the correlation between $r_A$ and $r_v$ (see figure \ref{figureS4}). We distinguish two cases: Particles that are trapped in the optical potential (blue dots) and untrapped particles traveling over the standing wave (red dots). 
For untrapped particles we underestimate the true velocity the more ($r_v<1$) the smaller the area ratio $r_A$. For a trapped particle (blue dots) the situation is different. We find that the estimated velocities are smaller than the true velocities ($r_v<1$) for an area ratio $r_A \lesssim 0.9$. In contrast to that, the true velocity is overestimated ($r_v>1$) for an area ratio  $1> r_A \gtrsim 0.9$.

We thus conclude: 
Both the analytical and the numerical considerations show that the area ratio is a good measure for the validity of our method to extract the final velocity. 
For the particle studied in the main text (figure \ref{figure3}) we obtain a ratio of $r_A=0.98(±0.06)$ (figure \ref{figureS3}) which assures that we do not underestimate the true final exit velocity with the value extracted at finite field (figure \ref{figure3}).
\subsection{Reconstruction of the particle trajectory}
We extract the particle position x(t) from the normalized scattering curve $S_N$ in the following manner: First we divide $S_N$ by a fitted Gaussian envelope and then we take the inverse of the sinusoidal part. We assume the local maxima to be positioned exactly at the antinodes. We distinguish two sorts of local minima: Those with a value near zero indicate that the particle passes through a node of the light field. Non-zero minima correspond to turning points of the particle while it is channelled along an antinode of the standing wave.
At both the maxima and the zero-valued minima the coupling is insensitive to the particle's change in position. This leads to discontinuities in the reconstruction of the trajectory, which are corrected with a linear regression.

\subsection{A second example of cavity cooling} 
\begin{figure*}

		\includegraphics[width=0.7\textwidth]{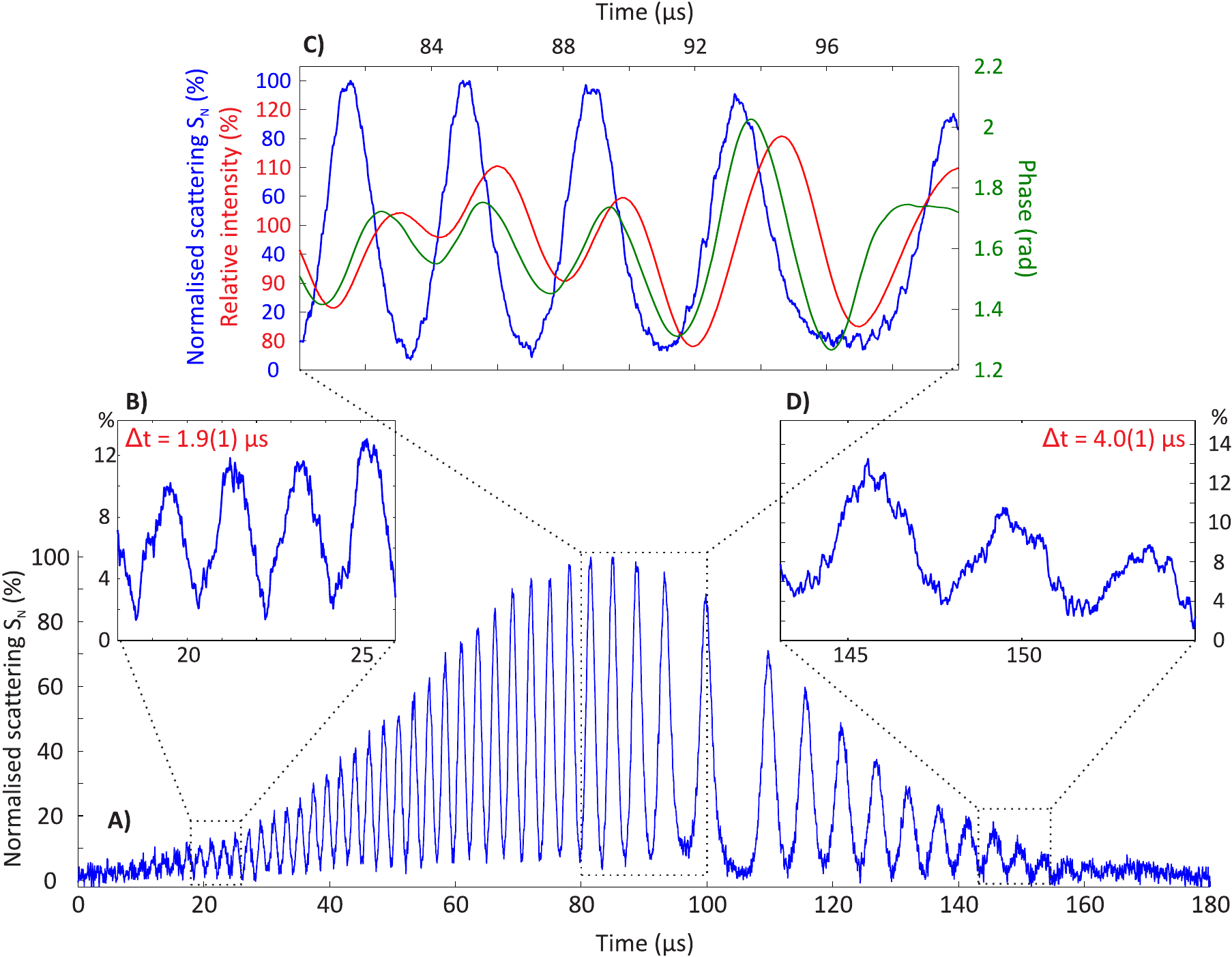}
\caption{$S_N (t)$, phase and intensity of the cavity field plotted for a second nanosilicon sample. Its initial and final transverse velocity $v_{x,in} =41(2)$ cm/s and $v_{x,out}  = 20(1)$ cm/s are derived from the modulation of $S_N$ in panels B and D. The phase and intensity of the cavity respond with the time delay that is required for cooling (panel C). }
\label{figureS5}
\end{figure*}
Figure \ref{figureS5} A) shows the normalised scattering signal $S_N (t)$ of a second nanoparticle in transit through the cavity. We find a forward velocity of $v_z=2.1$ m/s. The initial transverse velocity is $v_{x,in}=41(2)$ cm/s (panel B). 
While the particle travels through the mode, the phase and intensity of the cavity respond with the phase delays that are required to realize transverse cooling (see panel C). Contrary to the particle of figure 3, this second particle runs over the standing wave while leaving the cavity mode. Hence, the final transverse velocity can be measured directly as before and we obtain $v_{x,out}  = 20(1)$ cm/s. 
This results in a reduction of the transverse kinetic energy by a factor of $4.2(2)$. In comparison to the first particle we find a smaller cooling factor, which can be explained by the higher forward velocity and the shorter interaction time with the cavity mode.

\end{appendix}
\end{document}